\documentclass[a4paper,onecolumn, 11pt]{quantumarticle}
\pdfoutput=1
\usepackage[utf8]{inputenc}
\usepackage[english]{babel}
\usepackage[T1]{fontenc}
\usepackage{amsmath}
\usepackage{hyperref}

\usepackage{tikz}
\usepackage{lipsum}

\newcommand{\bra}[1]{\ensuremath{\langle{#1}|}}
\newcommand{\ket}[1]{\ensuremath{|{#1}\rangle}}

\begin{document}

\title{Nonlinear bosonic Maxwell's demon}

\author{Atirach Ritboon}
\affiliation{Research Unit in Energy Innovations and Modern Physics (EIMP), Thammasat University, Khlong Nueng, Khlong Luang, Pathum Thani, 12120, Thailand.}
\affiliation{Department of Optics, Palacky University, 17. listopadu 12, 771 46 Olomouc, Czech Republic.}
\email{ritboon@tu.ac.th}
\orcid{0000-0003-2519-748X}
\author{Radim Filip}
\email{filip@optics.upol.cz}
\orcid{0000-0003-4114-6068}
\affiliation{Department of Optics, Palacky University, 17. listopadu 12, 771 46 Olomouc, Czech Republic.}
\maketitle

\begin{abstract}
  Maxwell's demon principle of extracting valuable resources through measuring fluctuations in the system already stimulated modern quantum physics. In contrast to classical physics, a free coupling to a probe and its free measurement fundamentally shape the system state. This becomes a new dimension of the Maxwell demon effect, as in addition to the gained information, the back action on the system can be exploited and essential for further applications. We investigate quantum bosonic Maxwell's demon coupled to a two-level system to address this issue straightforwardly. The deterministic multiple subtractions of energy quanta by an energetically conservative Jaynes-Cummings interaction leads to an out-of-equilibrium state. Although still super-Poissonian, it can resonantly excite another two-level system  better than any thermal state. To further reduce the super-Poissonian statistics close to a Poissonian by a Maxwell's demon operation and increase the excitation rate, we suggest subsequent use of still energetically conservative multiphonon subtractions performed by an available nonlinear Jaynes-Cummings interaction. The optimal combination of both deterministic subtractions leads to statistics that approaches a Poissonian distribution otherwise produced by shot-noise-limited sources as an ideal laser requiring extreme bosonic nonlinear saturations.
\end{abstract}

\section{Introduction}
Maxwell's demon, a thought experiment, demonstrates that, if ones could access to the information about the state of a system through a classical measurement, then they can exploit such information to gain mechanical work or energy from the system through a classical control over it. This thought experiment leads to the generalization of the second law of thermodynamics by emphasizing the possibility of information-work conversion. It is one of the vital principles that rectify thermal fluctuations, without using strong nonlinearity, simply by measurement and classical control \cite{Maruyama2009}. In classical Maxwell's demon, the classical measurement is ideally arbitrarily precise without any back action on the system, as a measured quantity of a system is treated as a hidden variable. It changes dramatically in the quantum domain, as the different couplings to a probe and its subsequent measurement form new states of the system. Such events often turn out to be destructive; however, they sometimes can conditionally distill the system into a more useful resource \cite{Bennett1996, Kwiat2001}, as broadly explored in quantum information, especially in resource theories. If the distillation fails, it can still be repeated continuously until obtaining the resource. In this way, the chance of failure, in principle, can be minimized to zero at the cost of the speed or protocol multiplexing. The initial thermal fluctuations are used at the output with a negligible probability. Practically, it does not significantly influence the generated resource. The quantum transformation to a better resource, therefore, depends on the coupling and measurement back action, not only on information like in   an ideal classical measurement where any back action is left unconsidered \cite{Parrondo2015, Toyabe2010, Debiossac2020}, and it is regarded as the key point of this work.
First, we need to define free states, free couplings, and free measurements \cite{Chitambar2019, Brandao2013}. The free states are naturally thermal equilibrium states diagonal in the energy eigenbasis. The essential free unitary coupling of the system to the probe is then energy-conserving; the free probe energy measurement commutes with the probe's energy. Free controlled operations are also energy conserving couplings with an ancilla in thermal states that maximally have the energy of the input ones. It defines the most basic but nontrivial playground to explore and compare Maxwell demon methods mutually and also with other noise rectification strategies.  

The quantum Maxwell demon is more involved and diverse for a bosonic system representing a single mode of photons, phonons or other bosonic particles. Here, the simplest case of a free coupling is an energy-conserving beam-splitter type of resonant coupling. After this beam-splitter coupling, macroscopic measurement integrating energy already allows conditional manipulation with continuous energy statistics, for example, used in \cite{Iskhakov2016a, Iskhakov2016b}. Microscopic single-quanta detection opens space for subtracting individual energy quanta conditionally \cite{Parigi2007, Zavatta2008, Fedorov2015, Bogdanov2017, Katamadze2018, Katamadze2019, Katamadze2020, Enzian2021}, even for macroscopic thermal states, and charging the macroscopic battery by average energy \cite{Vidrighin2016, Shu2017}.  

However, for microscopic phononic states with few quanta on average, the statistics after subtraction becomes crucial for charging a microscopic battery. Such battery is represented by a two-level system coupled to the phonons, light or microwave fields. Multiple subtractions increase mean energy and reduce autocorrelation between quanta causing them to be more statistically independent \cite{Barnett2018, Bogdanov2017}. They mainly increase the mean-to-deviation ratio of the system’s energy, which is essential for information theory and thermodynamics \cite{Hlousek2017}. A deterministic, repeat-until-success version also deterministically excites a two-level system beyond any thermal state with stimulating thermodynamical consequences \cite{Hlousek2022}. Moreover, as recently demonstrated, the correlations between two thermal baths allow a Maxwell's demon based protocol to extract more work \cite{Zanin2022} and measurement strategies have been used in quantum memristors \cite{Spagnolo2022}.

In this work, we propose a nonlinear bosonic Maxwell's demon working at quantum level through simple and deterministic protocol, which is expected be straightforwardly realized in various quantum platforms. We first investigates this deterministic Maxwell demon method for a broadly feasible energy-conserving coupling, a linear Jaynes-Cummings coupling \cite{Leibfried2003}, probing a bosonic system sequentially by two-level systems to reach an out-of-equilibrium state. We then prove that the output state can excite another two-level system better than any thermal states. Differently from photonic Maxwell’s demon, we consider phononic systems represented, for example, by the extensively used mechanical modes of a single atom \cite{Chu2018} or, recently, a macroscopic oscillator \cite{Shore1993}. Alternatively, microwave superconducting experiments can also be considered for the experimental tests \cite{Fink2008}. In these cases, usually, the mean thermal occupation per mode can be much higher than that of thermal light sources. Despite the low dimension of the probes, the deterministic linear subtraction increases both the mean-to-deviation ratio of energy and the probability of exciting two-level systems higher than that from thermal states. It proves the power of such operations beyond a conventional Fock state lowering \cite{Um2016}. We further involve still energy-conserving nonlinear Jayness-Cumming coupling \cite{Sukumar1981, Singh1982, Villas-Boas2019} available at trapped-ion platforms \cite{Scully1997}, cavity quantum electrodynamics \cite{Brune1987} and superconducting circuits \cite{Garziano2015} mentioned above to perform a nonlinear subtraction of more quanta at once. Trilinear interactions, additionally, can also considered as alternative options \cite{Ding2017a, Ding2017b, Maslennikov2019, Ding2018}. Remarkably, we prove that optimally implementing nonlinear subtractions after the linear ones increase both the mean-to-deviation ratio of energy and the probability of excitation of atoms. It is the first example of using two-quanta processes to bring the statistics of phononic mode closer to a Poissonian, without any classical external drive and intense nonlinear saturation typical for such processes in laser \cite{Scully1997}. It proves that Maxwell’s demon, based on available nonlinear energy-conserving couplings, can open a new territory for quantum statistical and thermodynamical investigations.

We study the deterministic enhancement of population inversion of two-level systems coupled with harmonic oscillators via the well-known Jaynes-Cummings (JC) interaction. This improvement is originated from the successive change in the population distribution of the harmonic oscillators as a result of sequential subtractions of the oscillators' excitation. If the subtractions fail, we classically replace the oscillators with thermal oscillators or ones obtained from the previous round of subtraction to keep the overall procedure deterministic. The phonon distribution, thereby, is modified into a bell-shaped so that the end-product oscillator can be used as a resource for better exciting a two-level system. 

\section{Results}

\begin{figure*}[ht]
\includegraphics[width=\textwidth]{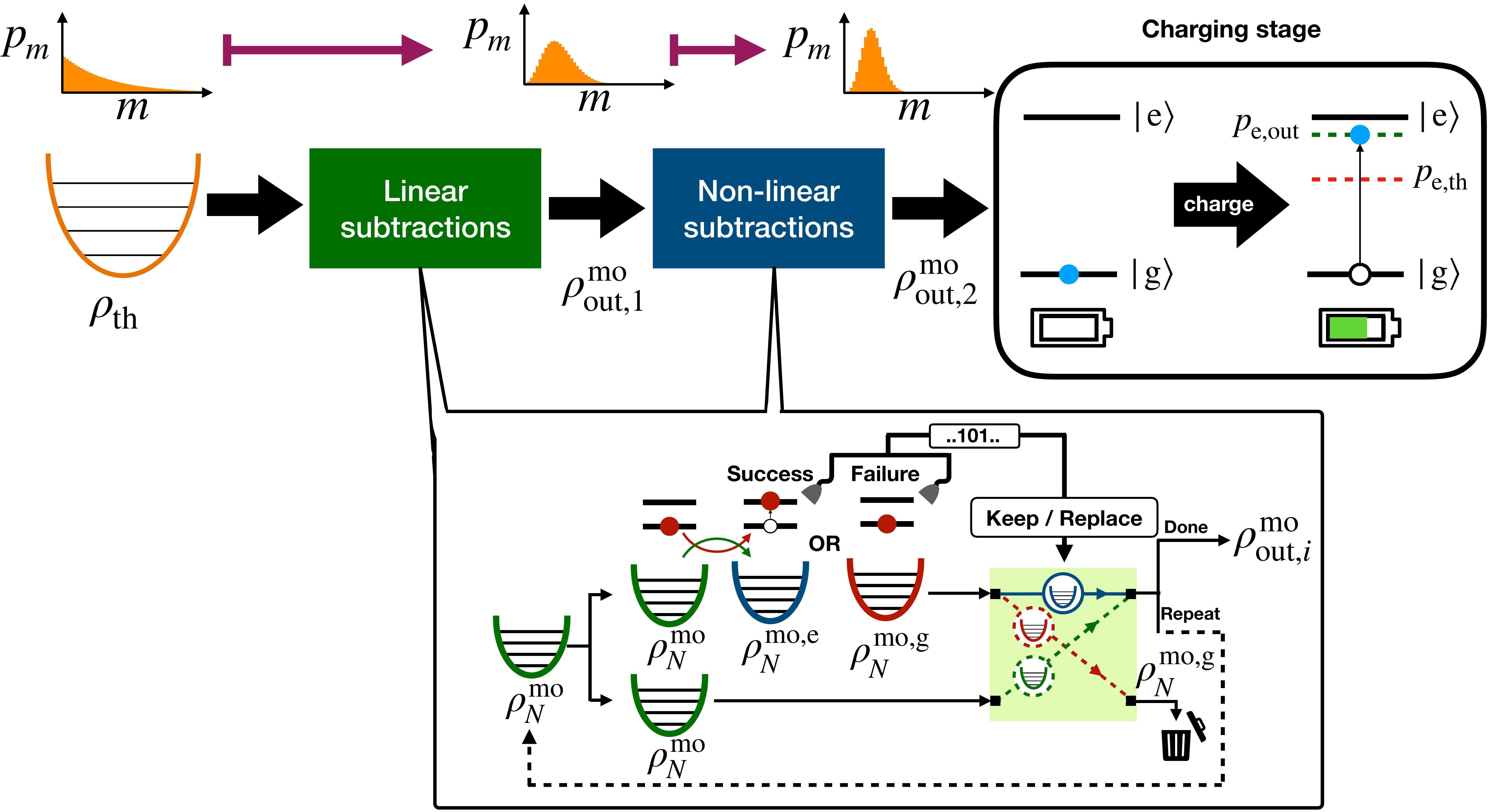}
\caption{The diagram represents the overview of the deterministic nonlinear bosonic protocol for resonantly charging a quantum battery, represented by a two-level system. The protocol is divided into two parts: phonon subtractions and the two-level charging process. The excitation subtraction also consists of two parts: linear (green) and nonlinear (blue) subtractions using the protocol II, illustrated in the bottom inset and described in detail in the later part of section \ref{sec:2}. Several linear subtractions are performed to gradually shape the probability distribution of a harmonic oscillator, initially in a thermal state $\rho_{\rm th}$, into a bell-shaped before nonlinear subtractions take place to trim and squeeze the probability distribution further. Such harmonic oscillators, hence, are used to charge the quantum batteries, initially being set to be in the ground state $\ket{\rm g}$, through Jaynes-Cummings interaction. The excitation probability $p_{\rm e, out}$ is expected to exceed its corresponding thermal bound.\label{fig:0}}
\end{figure*}
To understand the overall picture and procedure of this work, we devote this section to explain the overview of the proposed protocol employed to gradually shape the probability distribution of a harmonic oscillator into a bell-shaped for a better probability of exciting a two-level system, which is regarded as a quantum battery, via JC interaction. Figure \ref{fig:0} displays the overall processes of the scheme. A harmonic oscillator in thermal equilibrium with a thermal bath undergoes a linear excitation subtraction using a linear JC coupling for several times before further extracting its excitations through the nonlinear interaction so that its probability distribution becomes even more squeezed from both sides, at low and high quanta. The output harmonic oscillator is then used to charge a two-level battery through the linear JC coupling to examine if its performance exceeds the thermal bound, the maximum probability of exciting a qubit with a thermal state $\rho_{\rm th}$, the initial state it associated with.

The lower inset depict the excitation subtraction procedure of both linear and nonlinear subtractions. A harmonic oscillator in a motional state $\rho^{\rm mo}_N$ resonantly interacts with a two-level system through a JC interaction, with the interaction Hamiltonian linear in oscillator variables, or a nonlinear JC interaction \cite{Singh1982} having higher powers of the oscillator variables in the interaction Hamiltonian depending on which type of excitation subtraction being performed at that stage. After they interact for a strategically chosen period of time, the state of the qubit is measured. If the measurement results the excited state, the subtraction is thus performed successfully, otherwise it fails. The measured outcome then feeds forward to decide weather the harmonic oscillator could be kept or replaced by its previous successful version before repeating every steps again in the next round of subtraction.

\section{Linear subtraction}\label{sec:2}

\begin{figure*}[ht]
\centering
\includegraphics[width=0.7\textwidth]{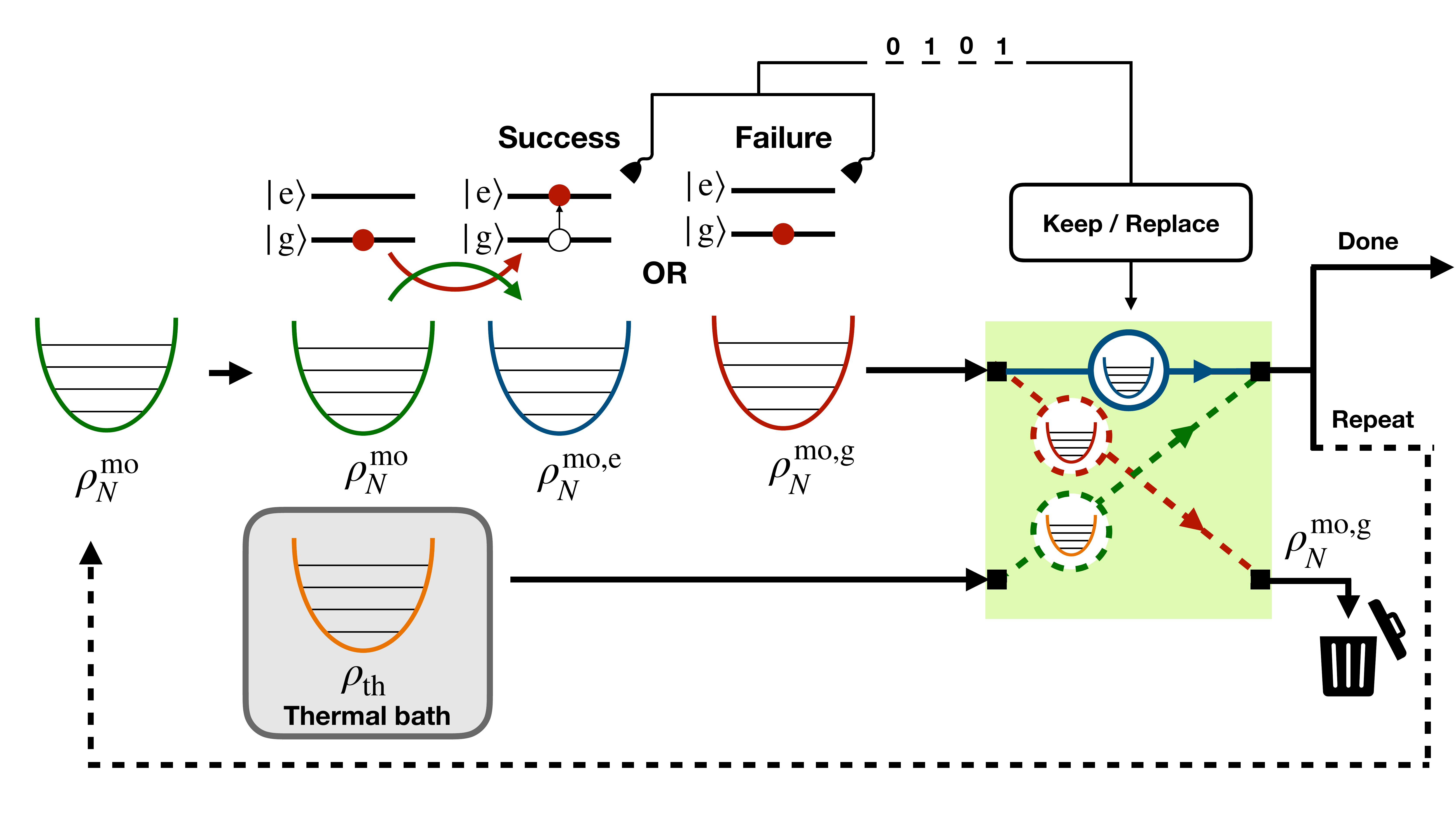}
\caption{Protocol I: the diagram illustrates the proposed phonon subtraction protocol. In the $N$th round of subtraction, an oscillator in a state $\rho^{\rm mo}_N$ couples with a two-level system at its ground state for the optimal interaction time $t^{\rm op}_N$. Then a measurement in the energy basis of the qubit is performed. If the excited state \ket{\rm e} is obtained, we keep the oscillator (blue), as its energy is successfully subtracted, for the next round, otherwise replacing it with a new oscillator in the initial thermal state.}\label{fig:1}
\end{figure*}
The linear subtraction is performed using a resonant JC interaction whose Hamiltonian in the interaction picture can be written as
\begin{align}\label{eq:2.1}
    \hat{H}_{\rm int} = \hbar\lambda\left(\hat{\sigma}_{+}\hat{a}+\hat{\sigma}_{-}\hat{a}^\dagger\right),
\end{align}
where $\lambda$ is the coupling strength, $\hat{\sigma}_{+}$ ($\hat{\sigma}_{-}$) is the rising (lowering) operator of the qubit, and $\hat{a}^\dagger$ ($\hat{a}$) is the creation (annihilation) operator of the oscillators. We then can express the unitary operator associated with this coupling running for an interval time $t$ as
\begin{align}\label{eq:2.2}
    \hat{U}_{\rm JC}(t) =& \ket{\rm e}\bra{\rm e} \cos(\lambda t\sqrt{\hat{n}+1})+\ket{\rm g}\bra{\rm g}\cos(\lambda t \sqrt{\hat{n}})\nonumber\\ 
    &\quad-{\rm i}\hat{\sigma}_{-}\hat{a}^\dagger\frac{\sin(\lambda t \sqrt{\hat{n}+1})}{\sqrt{\hat{n}+1}}
    -{\rm i}\hat{\sigma}_{+}\hat{a}\frac{\sin(\lambda t\sqrt{\hat{n}})}{\sqrt{\hat{n}}},
\end{align}
where $\ket{\rm g}\bra{\rm g}$ and $\ket{\rm e}\bra{\rm e}$ denote the projections on the ground and excited states of the qubits respectively, and $\hat{n}=\hat{a}^\dagger \hat{a}$ is the number operator of the harmonic oscillators. 

The first proposed protocol, denoted as protocol I, for linearly subtracting the motional excitations is schematically illustrated in figure \ref{fig:1}. At the beginning, harmonic oscillators are in thermal equilibrium with a thermal bath at temperature $T$, while the probes, two-level systems, are prepared in the ground state $\ket{\rm g}$. The initial composite state of them thus can be expressed as
\begin{align}\label{eq:2.3}
    \rho_{0} = \ket{\rm g}\bra{\rm g}\otimes\rho_{\rm th},
\end{align}
where $\rho_{\rm th}$ denotes the state of a harmonics oscillator in thermal equilibrium with a mean number of excitations $\bar{n}$,
\begin{align}\label{eq:thermal}
\rho_{\rm th}&=\sum^{\infty}_{m=0}\frac{\bar{n}^m}{(\bar{n}+1)^{n+1}}\ket{m}\bra{m}\nonumber\\
&\equiv  \rho^{\rm mo}_0,
\end{align}
and is also regarded as the initial motional state of the oscillators.

The mean number of motional excitations $\bar{n}$ is related to the temperature $T$ by $\bar{n}=(\exp(\hbar\omega/k_{\rm B}T)-1)^{-1}$, where $\omega$ is the angular frequency of the oscillators. The interaction between the qubits and the oscillators is run for the optimal time, $t^{\rm op}_0$, chosen to maximize the probability of exciting the qubits, approximately related to $\bar{n}$ as $\lambda t^{\rm op}_0\approx\pi/(2\sqrt{\bar{n}+1})$ (see the appendix for more details). Subsequently, the measurement in the eigenbasis $\{\ket{\rm g}, \ket{\rm e}\}$ on these qubits is performed. We then postselect only those oscillators with the probes in the excited state $\ket{\rm e}$ to be used in the further steps of the protocol. The measurement and postselection project the state of the qubits and the harmonic oscillators onto
\begin{align}\label{eq:2.4}
    \rho^\prime_1(t^{\rm op}_0) =& \frac{1}{P^{(0)}_{\rm e}(t^{\rm op}_0)}\left(\hat{\sigma}_{+}\hat{a}\frac{\sin(\lambda t^{\rm op}_0 \sqrt{\hat{n}})}{\sqrt{\hat{n}}}\rho_0\frac{\sin(\lambda t^{\rm op}_0 \sqrt{\hat{n}})}{\sqrt{\hat{n}}}\hat{a}^\dagger\hat{\sigma}_{-}\right)\nonumber\\
    \nonumber\\
    =&\ket{\rm e}\bra{\rm e}\otimes\sum^{\infty}_{m=0}\frac{\bar{n}^{m+1}}{(\bar{n}+1)^{m+2}}\frac{\sin^2(\lambda t^{\rm op}_0\sqrt{m+1})}{P^{(0)}_{\rm e}(t^{\rm op}_0)}\ket{m}\bra{m}\nonumber\\
    =&\ket{\rm e}\bra{\rm e}\otimes \rho^{\rm mo, e}_0,
\end{align}
where $P^{(0)}_{\rm e}(t^{\rm op}_0)$ is the probability of observing the excited state $\ket{\rm e}$ at the optimal time, which acts as the normalization factor of the term in the bracket and $\rho^{\rm e}_0$ is the  state of the oscillators after the postselection. The qubits of those postselected systems are then reset back to their ground state $\ket{\rm g}$ by the dissipation of their energy to the environment. We consider that the qubit dissipation is much slower than the JC interaction and, therefore, do not decohere the process. After that, we replace those failed systems with new systems in the initial state $\rho_0$. All mentioned processes are then repeated again, but this time the initial motional state of the ensemble for the new round has changed from $\rho^{\rm mo}_0$ due to the measurement back action of the first subtraction. For the $N$th round of repeat-until-success subtraction by the JC interaction, the state of the ensemble can be expressed as
\begin{align}\label{eq:2.5}
    \rho^{\rm mo}_N = \hat{a}\frac{\sin(\lambda t^{\rm op}_{N-1} \sqrt{\hat{n}})}{\sqrt{\hat{n}}}\rho^{\rm mo}_{N-1}\frac{\sin(\lambda t^{\rm op}_{N-1} \sqrt{\hat{n}})}{\sqrt{\hat{n}}}\hat{a}^\dagger + (1-P^{(N-1)}_{\rm e}(t^{\rm op}_{N-1}))\rho^{\rm mo}_0,
\end{align}
where $\rho^{\rm mo}_{N-1}$  is the achieved motional state of the previous subtraction,  $P^{(N-1)}_{\rm e}$ is the probability of getting the excited state $\ket{\rm e}$ in the previous round, and $t^{\rm op}_{N-1}$ is the interaction time that maximizes $P^{(N-1)}_{\rm e}$. Note that the motional state in Eq. \ref{eq:2.5} when $N=1$ differs from the state in Eq. \ref{eq:2.4} by additional terms, associated with the repeat-until-success subtractions as studied in \cite{Marek2018}. The probability $P^{(N)}_{\rm e}$ of getting excited state at the optimal interaction time $t^{\rm op}_{N}$ can be written as
\begin{align}
    P^{N}_{\rm e}(t^{\rm op}_{N}) = \sum^{\infty}_{m=0} p_m^{(N)}\sin^2(\lambda t^{\rm op}_{N} \sqrt{m}).
\end{align}
where the probability distribution $p_m^{(N)}$ of the harmonic oscillators in the state $\rho_N$ can be expressed as
\begin{align}\label{eq:2.6}
    p^{(N)}_m = p^{(N-1)}_{m+1}\sin^2(\lambda t^{\rm op}_{N-1}\sqrt{m+1})+(1-P^{(N-1)}_{\rm e}(t^{\rm op}_{N-1}))p^{(0)}_m,
\end{align}
 We denote $p^{(N-1)}_m$ to be the achieved probability distribution in the previous round with $p^{(0)}_m=\bar{n}^m/(\bar{n}+1)^{m+1}$, the initial probability distribution at thermal equilibrium. This equation describes how each subtraction gradually shapes the  probability distribution of the oscillators in each round. 

\begin{figure}[ht]
\centering
\includegraphics[width=0.9\linewidth]{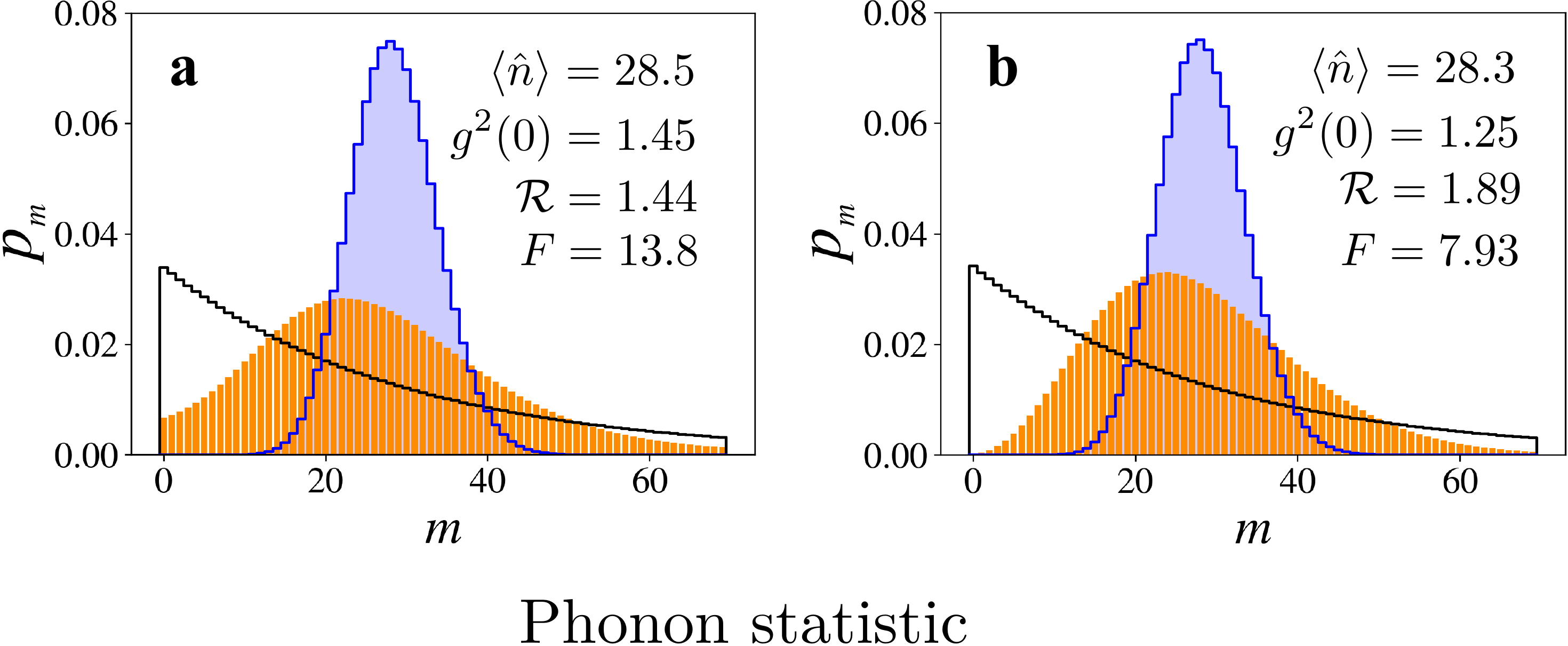}
\caption{The change in the population distributions after five linear subtractions using ({\bf a}) protocol I and ({\bf b}) protocol II is demonstrated. The black solid lines represent the initial population distribution of the oscillators being in thermal equilibrium with the mean excitation number $\bar{n}=30$. Sequential subtractions gradually forms a bell-shaped probability distributions (orange) with its peak located slightly lower than the initial mean excitation $\bar{n}=30$. The distribution obtained from protocol II, in figure \ref{fig:3}b, is noticeably narrower than that obtained from protocol I but has slightly lesser mean excitation, $\langle \hat{n}\rangle$. These two distributions are then compared to Poisson distributions (blue) with the same average phonon numbers. Their relevant information are also given, including their mean excitation $\langle \hat{n}\rangle$, their second-order correlation functions $g^2(0)$, their mean-to-deviation ratios of excitation $\mathcal{R}$, defined in Eq. \ref{eq:3.3} and their Fano factors $F$.}\label{fig:3}
\end{figure}

Let us consider the semiclassical case when the average excitation is very large, $\bar{n}\gg1$. The square of the sine function, the first term in Eq. (\ref{eq:2.6}), acts as a population filter. After the first subtraction, $N=1$, with the optimal time $\lambda t^{\rm op}_0\approx\pi/(2\sqrt{\bar{n}+1})$, the probabilities $p^{(1)}_n$ with the value of $n$ very different from the initial average excitation $\bar{n}$ are suppressed as the values of $\sin^2(\pi\sqrt{m+1}/(2\sqrt{\bar{n}+1}))$ is considerably smaller than unity, while those probabilities $p^{(1)}_m$ with $m$ close to $\bar{n}$ dominate the new probability distribution. This filtering effect still holds true for the subsequent subtractions, and it also makes the optimal interaction times of several further rounds are approximately the same as the first one: $\lambda t^{\rm op}_{N}\approx\lambda t^{\rm op}_{0}\approx \pi/(2\sqrt{\bar{n}+1})$ for $N\ll\bar{n}$. As the probability of having the qubit in the excited state  grows progressively with the number of performed subtractions, as shown in figure \ref{fig:4}, the last term in Eq. (\ref{eq:2.6}) gradually becomes a smaller contribution. 
Each subtraction of the motional excitation, with optimized coupling, thus gradually modifies the probability distribution $p^{(N)}_m$ into a bell-shaped centered around $\bar{n}$, as depicted in figure \ref{fig:3}. The center of the probability distribution is shifted to the left noticeably, when the number of subtractions becomes comparable with the initial average excitation number $\bar{n}$, indicating that the average excitation number decreases slightly each time we we perform a subtraction. 

\begin{figure*}[ht]
\centering
\includegraphics[width=0.7\textwidth]{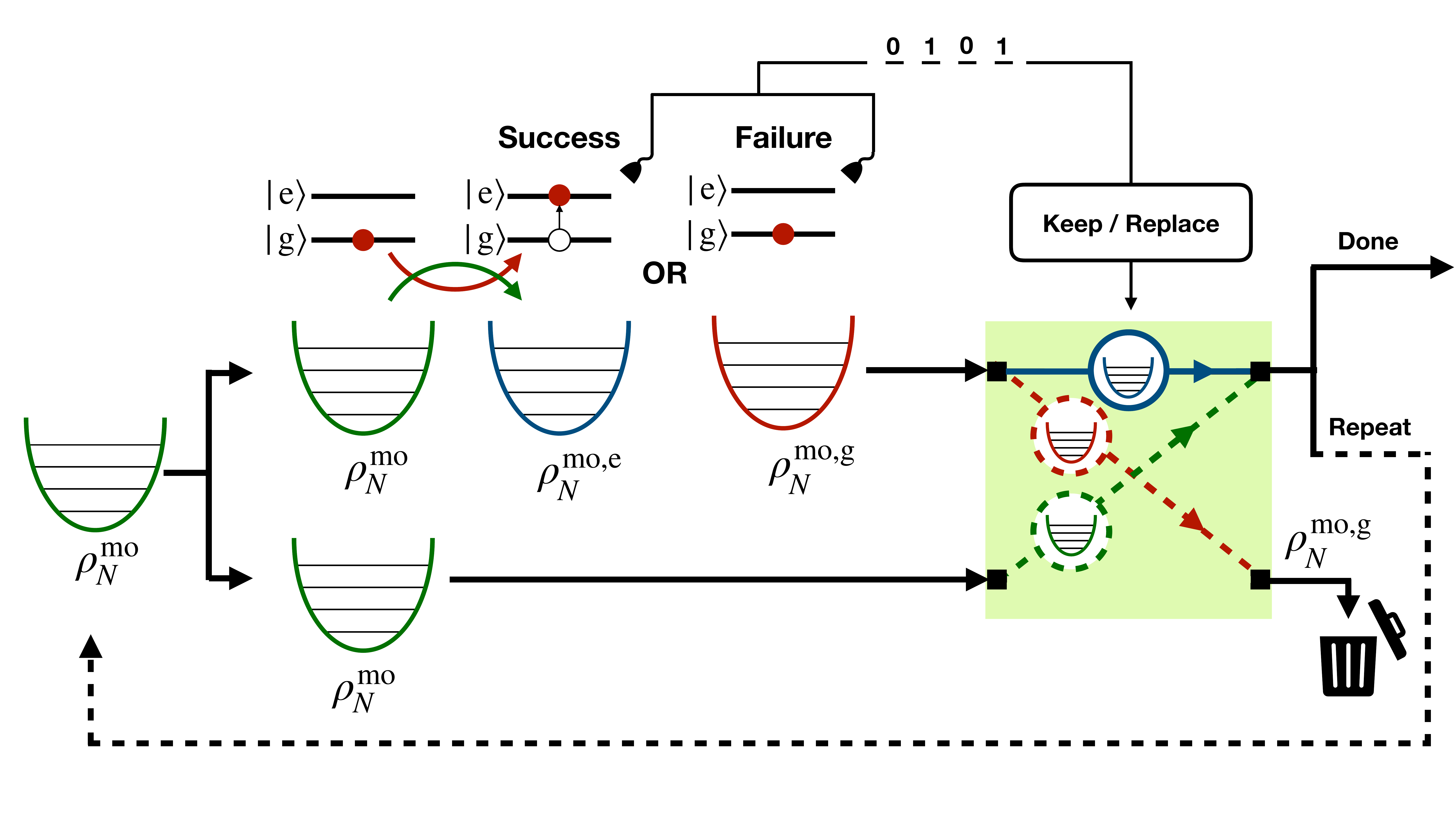}
\caption{Protocol II: the diagram of the improved protocol is similar to that of protocol I shown in figure \ref{fig:1} but in this case the failed systems are replaced by the successful systems obtained in the previous round instead of using systems in the initial thermal state.}\label{fig:2}
\end{figure*}

Protocol I can be further improved, if in the last step of each subtraction, instead of replacing the failed systems with systems in thermal equilibrium, $\rho^{\rm mo}_0$, we replace them with the successfully achieved systems of the previous round. The diagram of the second protocol, named protocol II, is depicted in figure \ref{fig:2}. The achieved state of the ensemble after the $N$th round of subtraction becomes
\begin{align}\label{eq:2.7}
    \rho^{\rm mo}_N = \hat{a}\frac{\sin(\lambda t^{\rm op}_{N-1} \sqrt{\hat{n}})}{\sqrt{\hat{n}}}\rho^{\rm mo}_{N-1}\frac{\sin(\lambda t^{\rm op}_{N-1} \sqrt{\hat{n}})}{\sqrt{\hat{n}}}\hat{a}^\dagger + (1-P^{(N-1)}_{\rm e}(t^{\rm op}_{N-1}))\rho^{\rm mo}_{N-1},
\end{align}
where the population distribution of the harmonic oscillators is modified as
\begin{align}\label{eq:2.8}
     p^{(N)}_m = p^{(N-1)}_{m+1}\sin^2(\lambda t^{\rm op}_{N-1}\sqrt{m+1})+(1-P^{(N-1)}_{\rm e}(t^{\rm op}_{N-1}))p^{(N-1)}_m.
\end{align}
This modification can suppress both tails of the bell-shaped probability distribution faster and better than protocol I at the cost of collecting and storing the outcomes of the previous steps. 

We assume that the thermalization time of the motional state is very long compared to the total time spent in all processes of protocols I and II, so that the heat transferred from the thermal bath to the considered oscillators is very small and negligible. The thermalization effect, as a result, can be 
ignored.
\section{Population inversion}\label{sec:3}
A population inversion of qubits happens when the probability $P_{\rm e}$ of finding the qubits in the excited state $\ket{\rm e}$ exceeds the probability $P_{\rm g}$ of finding them in the ground state $\ket{\rm g}$, \emph{i.e.} $P_{\rm e}-P_{\rm g}>0$ or $P_{\rm e}>1/2$.
We then devote this section to demonstrate and explain the performance of sequential linear subtractions. Let us first discuss the relation between the population distribution of a harmonic oscillator and the maximum excitation probability of a two-level system. For a qubit initially being in the ground state $\ket{\rm g}$, the probability $P_{\rm e}$ of getting the excited state $\ket{\rm e}$ after it is coupled an oscillator via the JC interaction for $t$ is
\begin{align}\label{eq:3.1}
    P_{\rm e}(t)=\sum^{\infty}_{m=0}p_m\sin^2(\lambda t\sqrt{m}),
\end{align}
where $p_m$ is the population distribution of the oscillator. As an oscillator being in its motional ground state $\ket{0}$ cannot be coupled with a qubit in the ground state $\ket{\rm g}$, the probability $P_{\rm e}$ thus must be smaller than $1-p_0$, where $p_0$ is the probability of finding the oscillator in its ground state. This means the desired probability distribution should have a small probability $p_0$. For a weak coupling case in which $\lambda t\ll 1$, the probability approaches a linear rule $(\lambda t)^2\langle \hat{n}\rangle$ and the statistics of the oscillator do not matter in the classical excitation limit. For stronger coupling, however, this simple approximation breaks. Each term in the summation oscillates in time with different frequency depending on its index $m$. Narrower probability distributions thus give constructive interference of the oscillations, as they cause smaller mis-match between the oscillating frequencies, $\lambda \sqrt{m}$, of the dominant probabilities $p_m$ and provide a higher chance of getting the excited state $\ket{\rm e}$. For example, the perfect scenario in which the excited state $\ket{\rm e}$ is obtained via the JC interaction for certain is when the oscillator is in an arbitrary Fock state $\ket{n}$ as we can just choose the interaction time $t$ precisely to match $\lambda t \sqrt{n}=\pi/2$, leading to $P_{\rm e}=1$. Another factor to be considered for a bell-shaped distribution, with probabilities $p_m$ falling shapely when being far from the peak, is the mean excitation number, $\langle\hat{n}\rangle$. If two distributions have an identical bell-like shape but different mean excitation numbers, the one with a larger mean excitation can give larger $P_{\rm e}$. A larger $\langle\hat{n}\rangle$ provides smaller mis-match between the oscillating frequencies $\lambda \sqrt{m}$ of the terms in the summation. When $m$ differs from the mean excitation number $\langle\hat{n}\rangle$ by $\Delta m$ such that $m=\langle\hat{n}\rangle+\Delta m$ and $|\Delta m |\ll \langle\hat{n}\rangle$, the oscillating frequency of the probability $p_m$ can be approximated as
\begin{align}\label{eq:3.2}
\lambda \sqrt{m}&=\lambda \sqrt{\langle\hat{n}\rangle+\Delta m}\nonumber\\
&\approx\lambda \sqrt{\langle\hat{n}\rangle}+\frac{\lambda \Delta m}{2 \sqrt{\langle\hat{n}\rangle}}.
\end{align} 
The difference between the oscillating frequencies of the dominant probabilities in the summation of Eq. \ref{eq:3.1} is, therefore, inversely proportional to $\sqrt{\langle\hat{n}\rangle}$. Of course, the mean excitation becomes irrelevant when it comes to the case of an excited Fock state, as demonstrated earlier that, with a single oscillating term in the summation, $P_{\rm e}=1$ can be obtained for certain regardless of the mean excitation number. However, we need to bear in mind that the statistics will immediately play a crucial role once there exists small deviation from Fock states. From these discussed facts, among the parameters commonly used for analyzing the statistics of excitation, such as second-order correlation function, $g^2(0)=\langle \hat{a}^{\dagger 2}\hat{a}^2\rangle/\langle\hat{a}^\dagger\hat{a}\rangle^2$, and Fano factors $F=\langle (\Delta \hat{n})^2\rangle/\langle\hat{n}\rangle$, the appropriate parameter indicating the desirable phonon statistics, motivated by Eq. \ref{eq:3.2}, should be the mean-to-deviation ratio (MDR) of the population, denoted by $\mathcal{R}$, which is defined as
\begin{align}\label{eq:3.3}
    \mathcal{R}=\frac{\langle\hat{n}\rangle}{\sqrt{\langle (\Delta \hat{n})^2\rangle}}
\end{align}
where $\langle (\Delta \hat{n})^2\rangle=\langle\hat{n}^2\rangle-\langle\hat{n}\rangle^2$ represents the phonon fluctuation. It is more likely that the atom is excited better by a phonon with a greater value of $\mathcal{R}$.

\begin{figure}[ht]
\centering
\includegraphics[width=0.9\linewidth]{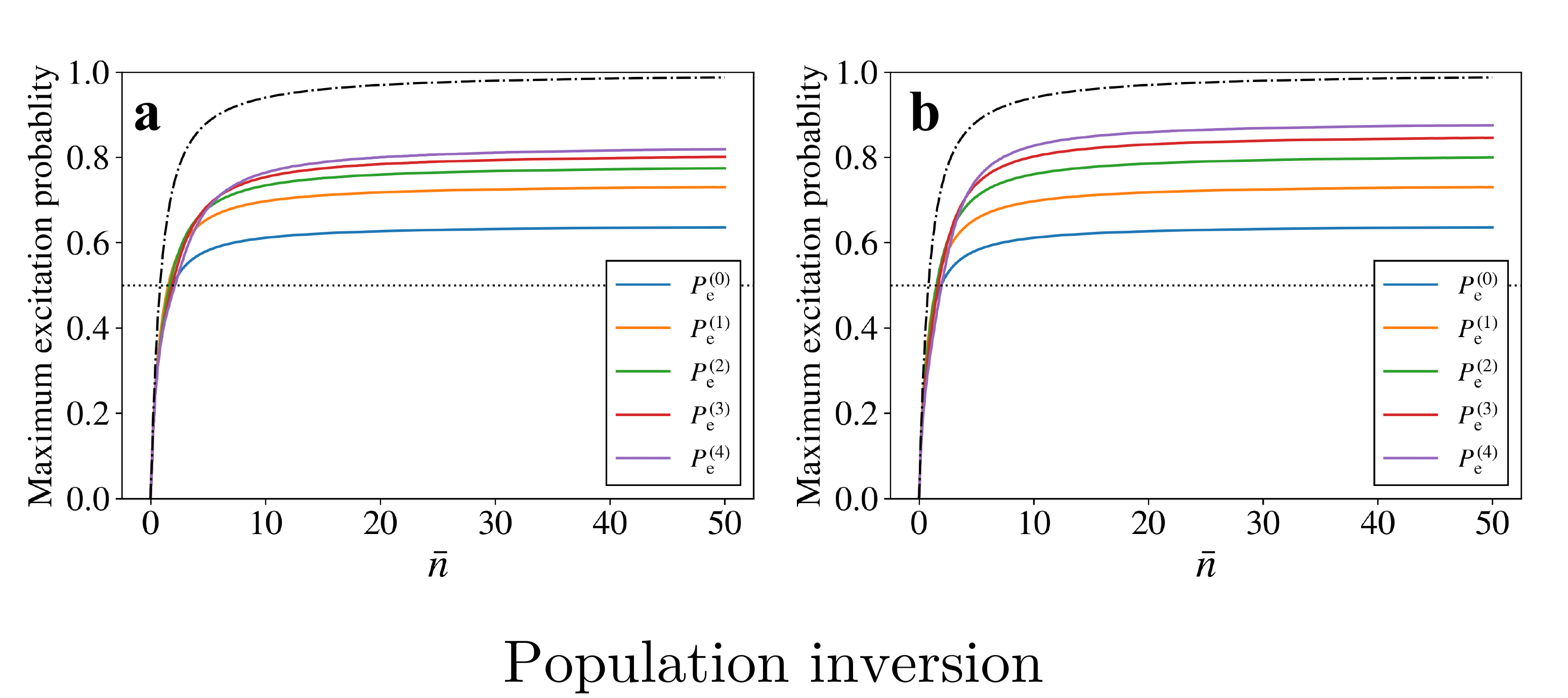}
\caption{The figure demonstrates the increase of the maximum probability $P^{(N)}_{\rm e}$ of getting the excited state $\ket{\rm e}$ through the JC coupling after the $N$th subtraction, when using ({\bf a}) protocol I, depicted in figure \ref{fig:1}, and ({\bf b})  protocol II, shown in figure \ref{fig:2}. The probability $P^{(N)}_{\rm e}$ increases sharply with the initial mean excitation $\bar{n}$ and reaches its plateau around $\bar{n}=5$. Protocol II apparently provides a better chance of exciting a two level-system compared to protocol I. The dotted lines are set to be at 0.5 to mark population inversion $P^{(N)}_{\rm e}>0.5$, while the dash-dotted lines display the excitation probability if the oscillator was in a coherent state. A thermal state with a very large mean excitation, $\bar{n}\gg 1$, can excites a qubit with probability of $P_{\rm e}\approx 0.6411$.}\label{fig:4}
\end{figure}

Figure \ref{fig:4} shows the maximum probability of having the excited state $\ket{\rm e}$ increases after each linear subtraction. This increase originates from the fact that the population distribution becomes narrower after each subtraction. The population distribution after five subtractions is shown in figure \ref{fig:3}. However, for a small initial mean excitation number, around $\bar{n}\sim 1$-$2$, the subtractions do not always increase $P^{(N)}_{\rm e}$ as the excitations of the oscillators are almost exhausted, {\it i.e.} most oscillators are in their motional ground state and no longer coupled with the two-level systems. As expected, since protocol II gives a smaller probability of being in the motional ground state, $p_0$, and a narrower probability distribution, it then gives higher probabilities $P^{(N)}_{\rm e}$, for $N\geq 2$. From the figure, we can clearly see that the increase of $P^{(N)}_{\rm e}$ gradually becomes saturated, as the value of $\Delta P^{(N)}_{\rm e}=P^{(N)}_{\rm e}-P^{(N-1)}_{\rm e}$ becomes smaller. Further subtractions barely increase the excitation probability. It is obvious from the figure that the saturated value of $P^{(N)}_{\rm e}$ obtained from protocol II is slightly greater than that from protocol I. This is because protocol II shapes the distribution in such a way that the probability $p_0$ and its neighborhood become very small, as shown in figure \ref{fig:3}b, compared to the distribution obtained from protocol I.

\section{Nonlinear subtraction}\label{sec:5}
As we pointed out in figure \ref{fig:3}, the distribution still has a long decaying tail for larger populations resembling the thermal statistics. To remove this limitation and shape the population distribution even faster and better, linear subtraction alone is no longer sufficient. From figure \ref{fig:4}, the performance of the linear subtractions eventually will reach its saturation, but there is still a way to break through it by utilizing a nonlinear interaction whose interaction Hamiltonian is of the form,
\begin{align}\label{eq:4.1}
    \hat{H}_{\rm non} = \hbar\lambda'\left(\hat{\sigma}_{+}\hat{a}^2+\hat{\sigma}_{-}\left(\hat{a}^\dagger\right)^2\right),
\end{align}
where $\lambda'$ denotes the coupling strength of the interaction. This interaction Hamiltonian would have the same form as the Hamiltonian in Eq. (\ref{eq:2.1}) if the annihilation and creation operators in Eq. (\ref{eq:2.1}) were replaced by their squares, $\hat{a}\rightarrow\hat{a}^2$ and $\hat{a}^\dagger\rightarrow\left(\hat{a}^\dagger\right)^2$. The unitary operator describing the time evolution of this nonlinear coupling is given as
\begin{align}\label{eq:4.2}
    \hat{U}_{\rm non}(t)=&\ket{\rm e}\bra{\rm e}\cos\left(\lambda' t \sqrt{(\hat{n}+1)(\hat{n}+2)}\right)+\ket{\rm g}\bra{\rm g}\cos\left(\lambda' t \sqrt{\hat{n}(\hat{n}-1)}\right)\nonumber\\
    \quad 
    &\quad-{\rm i}\sigma_{-}\left(\hat{a}^\dagger\right)^2\frac{\sin\left(\lambda't\sqrt{(\hat{n}+1)(\hat{n}+2)}\right)}{\sqrt{(\hat{n}+1)(\hat{n}+2)}} -{\rm i}\sigma_{+}\hat{a}^2\frac{\sin\left(\lambda' t\sqrt{\hat{n}(\hat{n}-1)}\right)}{\sqrt{\hat{n}(\hat{n}-1)}}.
\end{align}

As mentioned, nonlinear repeat-until-success subtractions are performed in the same way as linear subtractions explained in section \ref{sec:2}, except this time the employed interaction becomes nonlinear. However, unlike the linear case, a nonlinear subtraction cannot trim the tails of the population distribution desirably, see Appendix for more details. Therefore, the prior population distribution should be in a bell-shaped to some extend, and the probabilities $p_m$ associated with high-energy levels, $m\gg \bar{n}$, must be already very small. Otherwise, the nonlinear subtraction will form a ripple in the probability distribution of the harmonic oscillators, which is an undesired effect. Due to normalization condition, the probability distribution with a ripple is more dispersed compared to those without it. The probability $P_{\rm e}$, as a result, is not as large as it potentially should be.

\begin{figure}[h!]
\centering
\includegraphics[width=\textwidth]{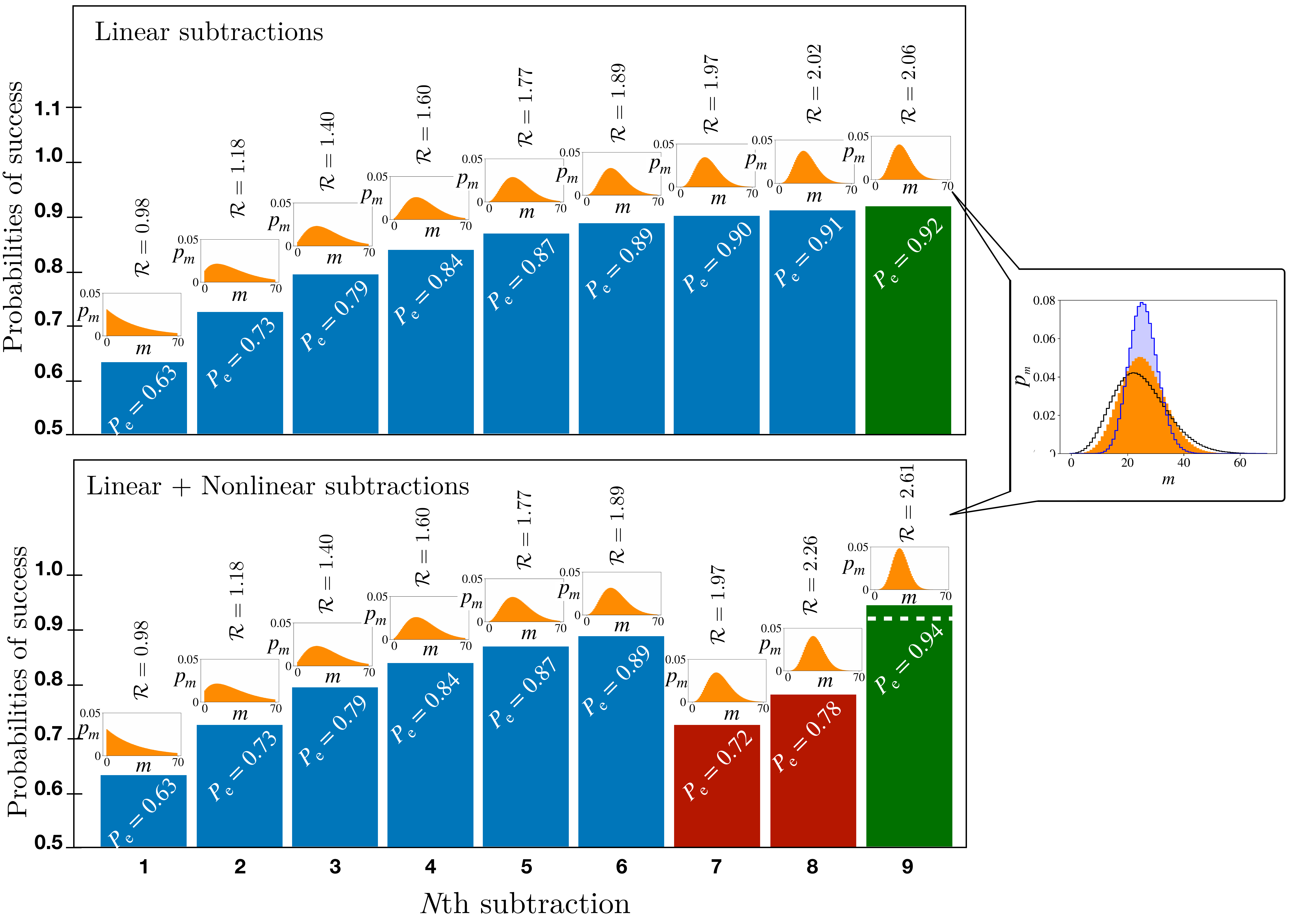}
\caption{The upper bar chart represents the probabilities of success,  obtaining the excited state, in $N$th subtractions using protocol II with an initial average phonon number of $\bar{n}=30$. On the other hand, the lower bar chart also shown such probabilities, but this time the 7th and 8th subtractions are performed by two consecutive nonlinear subtractions, with their probabilities of success denoted with the red bar. The actual probabilities of success, $P_{\rm e}$, are written in white on these bars so that we can see their fractional differences. Above each bar, a probability distribution of the harmonic oscillator contributing to the probability of success is displayed together with the value $\mathcal{R}$ of the mean-to-deviation ratio (MRD) of phonons. The last green bars of both bar charts represent the probabilities of getting the excited state through the JC coupling after eight subtractions, which can be regarded as the charging performance. To compare the performance of both schemes, we mark the height of the upper green bar on the lower one with the white dashed line. In the inset, the probability distributions of the two schemes of subtraction are compared. The solid black line represents the distribution obtained from eight linear subtraction while the orange histogram shows the distribution obtained when the two nonlinear subtractions are introduced. The later is then compared with a Poissonian distribution of the same mean excitation $\langle\hat{n}\rangle$, displayed by the blue histogram.}\label{fig:5}
\end{figure}

Consequently, in order to use nonlinear subtractions properly, we have to perform several linear subtractions using protocol II, displayed in figure \ref{fig:2}, so that the population distribution is modified into a proper bell-shaped with sufficiently short tails. After that, nonlinear subtractions using the procedure of protocol II can then be performed. The achieved state after a nonlinear subtraction can be expressed as
\begin{align}\label{eq:4.3}
\rho^{\rm f} = \hat{a}^2 S(\hat{n})\rho^{\rm i}S(\hat{n})\left(\hat{a}^\dagger\right)^2+\left(1-P^{\prime}_{\rm e}(\tau^{\rm op, })\right)\rho^{\rm i},
\end{align}
with
\begin{align}
S(\hat{n})=\frac{\sin\left(\lambda^\prime \tau^{\rm op} \sqrt{\hat{n}(\hat{n}-1)}\right)}{\sqrt{\hat{n}(\hat{n}-1)}}
\end{align}
where $\rho^{\rm i}$ is the former state of the oscillators before the nonlinear subtraction, $\tau^{\rm op}$ is the interaction time that gives the first locally optimal probability of successful subtraction. The probability of successful subtraction, on the other hand, reads
\begin{align}\label{eq:4.4}
P^{\prime}_{\rm e}(t)=\sum^{\infty}_{m=2}p^{\rm i}_m\sin^2\left(\lambda^{\prime} t\sqrt{m(m-1)}\right),
\end{align}where $p^{\rm i}_m$ is the prior probability distribution of the harmonic oscillators, the diagonal elements of $\rho^{\rm i}$.  In contrast to linear subtractions, the suitably chosen interaction time for nonlinear subtractions is not the time that optimizes the probability $P^{\prime}_{\rm e}$. From the equation, it is easy to notice that the optimal interaction time is approximately at $\lambda^\prime t =\pi/2$, resulting in $\sin^2\left(\lambda^{\prime} t\sqrt{m(m-1)}\right)\approx 1$ for $m>1$. However, with this interaction time, a nonlinear subtraction almost does not modified the probability distribution at all as the sine function barely shows its influence. We instead need to choose the interaction time $\tau^{\rm op}$  that gives the first locally optimal $P^{\prime}_{\rm e}$ such that $\lambda^\prime \tau^{\rm op}<\pi/2$. The interaction time $\tau^{\rm op}$ for this case is approximately related to the mean phonon number $2\langle \hat{n}\rangle$ as 
\begin{align}\label{eq:4.5}
    \lambda' \tau^{\rm op} \approx \frac{\pi}{2\langle \hat{n}\rangle}.
\end{align}
This means we can run a nonlinear interaction even faster than a linear one.
A nonlinear subtraction using protocol II manipulates the probability distribution as
\begin{align}\label{eq:4.6}
p^{\rm f}_m=p^{\rm i}_{m+2}\sin^2\left(\lambda^{\prime} \tau^{\rm op}\sqrt{(m+2)(m+1)}\right)+(1-P^{\prime}_{\rm e}(\tau^{\rm op}))p^{\rm i}_{m}.
\end{align}
Figure \ref{fig:5} compares the probabilities of success and their change after each subtraction of two different schemes: the scheme that employ only linear subtractions (the upper bar chart) and the scheme that use linear subtractions followed by nonlinear subtractions (the lower bar chart) to boost the performance before charging the quantum battery. From the upper bar chart, when only linear subtractions being used, the sequential increase of the probability of success gradually reaches its saturation. Therefore, additional linear subtractions just barely improves the probability of success and the charging performance, denoted by the last green bar.  The lower bar chart, on the other hand, demonstrates that the saturated performance can be further boosted with the help of nonlinear subtractions, whose probabilities of success are represented by the two red bars. The success probability of a nonlinear subtraction is noticeably lower than that of the previous linear subtractions, but it is still sufficiently large to make the nonlinear subtraction protocol practical. After six linear and two nonlinear subtractions, the charging performance or the probability of getting the excited state through the Jaynes-Cumming coupling, denoted by the green bar, becomes even larger than its previous version when all eight subtractions are linear. This increase lies in the change in the shape of the population distribution after the nonlinear subtractions. As shown in the inset of figure \ref{fig:5}, the distribution becomes more squeezed which is more desirable for exciting a two-level system.

We note here that a nonlinear subtraction cannot be used with the procedure of protocol I, as at the end of the protocol the failed systems are replaced by systems in a thermal state, which makes high-energy populations not sufficiently small. As a result, nonlinear subtractions cause the population distribution to be even more dispersed and a small ripple in the distribution to form.
\section{Discussion}
\begin{figure}[h!]
\centering
\includegraphics[width=0.6\linewidth]{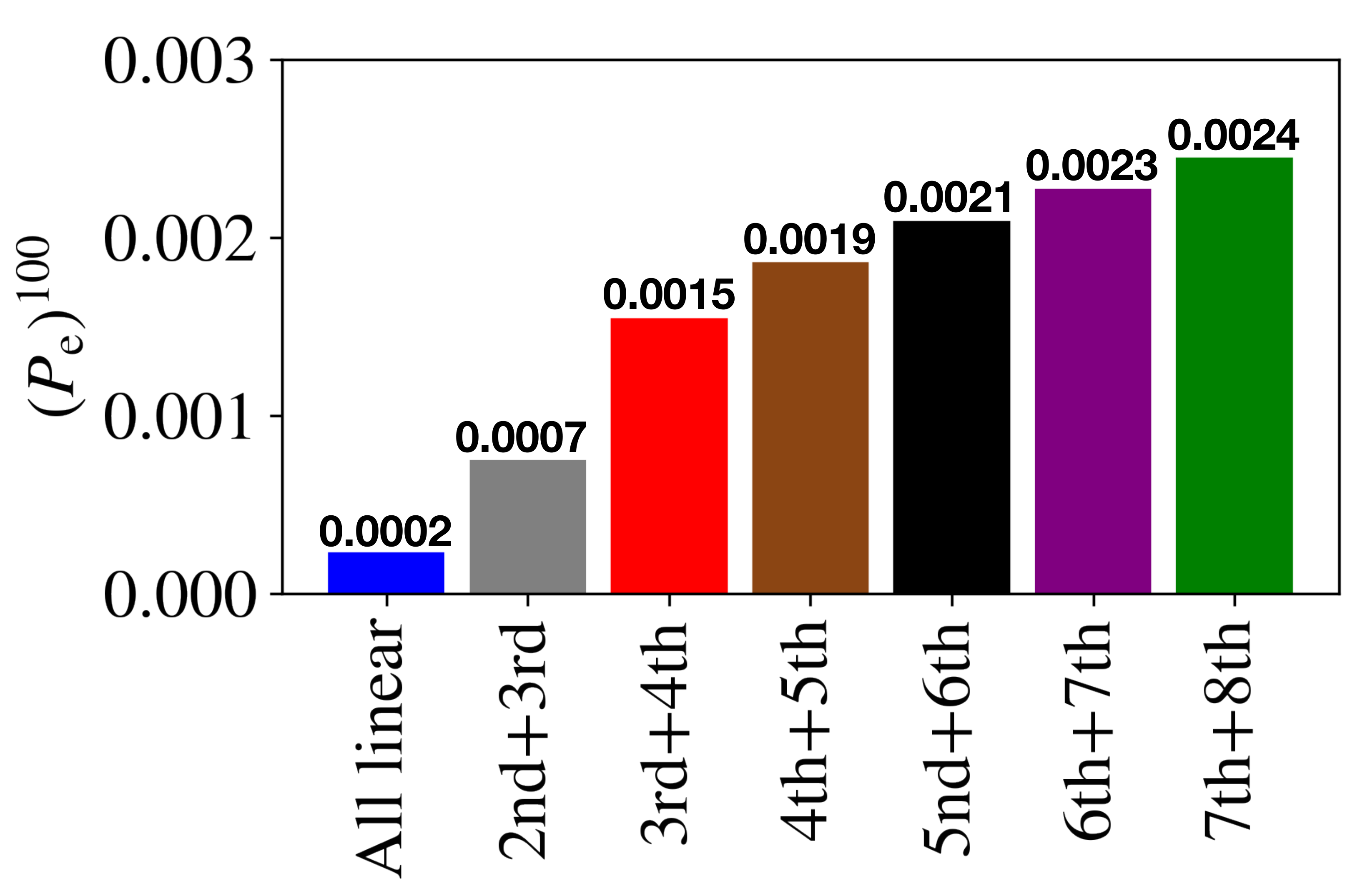}
\caption{The bar chart illustrates the probabilities $(P_{\rm e})^{100}$ of a hundred out of a hundred quantum batteries being successfully charged after eight phonon subtractions have been performed on an oscillator in a thermal state with $\bar{n}=30$ using different strategies including all eight subtractions being linear (the blue bar) and different combinations of six linear and two consecutive nonlinear subtractions. The labels $n$th+$(n+1)$th identify nonlinear subtractions in the sequence of subtractions for each strategy. It is apparent that the later nonlinear subtractions being performed the better the probability of success we achieve. The numbers above these bars are the actual values of such probabilities $(P_{\rm e})^{100}$.}\label{fig:6}
\end{figure}
The previous section compared the charging performance obtained from the two subtraction strategies: the linear-subtractions-only strategy and the combination of linear and nonlinear subtractions. It emerged that the later provides a better charging performance compared to the first. The remaining question is whether should nonlinear subtractions be performed at an earlier stage or at the very end of a subtraction sequence, or even something in between, to get the optimal charging performance. To answer this, we then examined different combinations of linear and nonlinear subtractions, where the coupling is optimized for each subtraction depending on the previous measurement outcomes. The result turns out to be that the later the nonlinear subtractions take place the greater the performance can be. The probabilities of successful charging an individual battery associated with these strategies of subtraction may, at first, look insignificantly different, but at the scale of mass production the differences become eventually magnified. This fact is demonstrated in figure \ref{fig:6} by comparing the probabilities of successfully charging a hundred out of a hundred quantum batteries using different combinations of linear and nonlinear subtractions. To relate it with the result in section \ref{sec:5}, we consider only the cases in which eight subtractions, including six linear and two consecutive nonlinear subtractions, are performed before the charging stage and compare their charging performances to that of the linear-subtractions-only strategy. From the figure, the best performance is obtained if the last two subtractions are nonlinear. Remarkably, its increase is even several time larger than the order of the excitation probability $(P_{\rm e})^{100}$ obtained through the only-linear-subtraction case. The underlying reason originates from the fact that nonlinear subtractions are better at squeezing the phonon population but improper for trimming its tails. Earlier use of nonlinear subtractions, as a result, reduces the mean phonon number of the end-product motional state and leads to a lower mean-to-deviation ratio $\mathcal{R}$. Several linear subtractions thus prepare a properly trimmed phonon distribution to be squeezed by the following nonlinear subtractions. 

\section{Conclusions}
The idea of classical Maxwell’s demon initiates the reformation  of the classical thermodynamics and generates a connection between information and thermodynamic work, and provides the fundamental idea of the conversion between these two quantities. With the information of a system obtained through measurements and precise control, the system then can be manipulated into an out-of-equilibrium state, and, in return, its energy can be extracted. The idea of such conversion is carried on to its quantum version with some fundamental differences. In contrast to the classical case, in which a measurement is treated to be arbitrarily sharp without any back action on the measured system, in the quantum domain, both measurement outcomes and their back actions unavoidably affect the way we conditionally control and manipulate the system in order to generate a useful resource. Each performed measurement not only extracts the system’s information but also transforms its state accordingly.

We have proposed a simple but deterministic protocol to realize a bosonic Maxwell’s demon at a quantum level exploiting the free coupling between the mechanical modes of a single atom and its internal electronic state. A measurement on the qubit with its outcome implying absorption of phonons by the qubit is regarded as phonon subtraction.  It is shown that linear subtractions from both  protocols I and II transforms the phonon state from an initial thermal state into an out-of-equilibrium state with a bell-shaped phonon distribution. This transformed motional state can eventually be used to charge a microscopic battery, another qubit, by exciting it through a linear JC coupling. The charging performance of such out-of-equilibrium states is higher than that of its initial thermal state, which can be  indicated by its increased mean-to-deviation ratio, $\mathcal{R}$. The performance is enhanced each time a subtraction being performed but becomes saturated eventually. To break through this limitation, a nonlinear subtraction, using a nonlinear JC coupling to absorb more phonons at once, must be exploited using the procedure of subtraction protocol II. The nonlinear interaction can boost the charging performance further, at the cost of its speed in the repeat-until-success protocols. It can further squeeze the phonon distribution better than the linear version, which increases $\mathcal{R}$ as a result. Nonetheless, it still has a drawback as it cannot trim the tails of the phonon population properly, making it better used as a final performance booster. Although, we use a trapped ion as an example quantum platform, this proposed protocol can also be realized easily in other platforms in which a nonlinear JC coupling is available, such as superconducting circuits and cavity quantum electrodynamics.

Optimized nonlinear subtractions after several linear subtractions can shape the end-product phonon statistics closer to, but not yet reach, a Poissonian without the help of an external classical drive. We, therefore, believe that nonlinear-based Maxwell’s demon can potentially pave the way for a new area of theoretical and experimental research in quantum statistics and thermodynamics.

\section{Author contributions}
R.F. conceived and supervise the project. A.R. and R.F. jointly build theory and analyze the results. A.R. performed numerical simulations and wrote the paper with feedback from R.F..

\begin{acknowledgements}
A.R. acknowledges funding support from the NSRF via the Program Management Unit for Human Resources \& Institutional Development, Research and Innovation (grant number B05F650024).
R.F. acknowledges support of project 22-27431S of the Czech Science Foundation.  The research was further supported by Thammasat University Research Unit in Energy Innovations and Modern Physics (EIMP) and also the European Union's 2020 research and innovation programme (CSA-Coordination and support action, H2020-WIDESPREAD-2020-5) under grant agreement No. 951737 (NONGAUSS). 
\end{acknowledgements}

\bibliographystyle{plain}

\begin{thebibliography}{35}
\bibitem{Maruyama2009} K. Maruyama, F. Nori, and V. Vedral, Colloquium
: The physics of Maxwell's demon and infomation, {\it Rev. Mod. Phys.} {\bf 81}, 1 (2009).

\bibitem{Bennett1996} C.H. Bennett, H.J. Bernstein, S. Popescu and B. Schumacher, Concentrating partial entanglement by local operations. {\it Phys. Rev. A} {\bf 53}, 2046–2052 (1996).

\bibitem{Kwiat2001} P.G. Kwiat, S. Barraza-Lopez, A. Stefanov and N. Gisin, Experimental entanglement distillation and ‘hidden’ non-locality, {\it Nature} {\bf 409}, 1014 (2001).

\bibitem{Parrondo2015} J. Parrondo, J. Horowitz, and T. Sagawa, Thermodynamics of information, {\it Nature Phys}. {\bf 11}, 131 (2015).

\bibitem{Toyabe2010} S. Toyabe, T. Sagawa, M. Ueda, E. Muneyuki, and M. Sano, Experimental demonstration of information-to-energy conversion and validation of the generalized Jarzynski equality, {\it Nature Phys.} {\bf 6}, 988 (2010).

\bibitem{Debiossac2020} M. Debiossac, D. Grass, J. J. Alonso, E. Lutz, and N. Kiesel, Thermodynamics of continuous non-Markovian feedback control, {\it Nature Comm.} {\bf 11}, 1360 (2020).

\bibitem{Chitambar2019} E. Chitambar and G. Gour, Quantum resource theories, {\it Rev. Mod. Phys.} {\bf 91}, 025001 (2019).

\bibitem{Brandao2013} F.G.S.L. Brand{\~{a}}o, M. Horodecki, J. Oppenheim, J.M. Renes, and R.W. Spekkens, Resource Theory of Quantum States Out of Thermal Equilibrium, {\it Phys. Rev. Lett.} {\bf 111}, 250404 (2013).

\bibitem{Iskhakov2016a} T. Sh. Iskhakov, V. C. Usenko, U. L. Andersen, R. Filip, M. V. Chekhova, and G. Leuchs, Heralded source of bright multi-mode mesoscopic sub-Poissonian light, {\it Optics Letters} {\bf 41}, 2149-2152 (2016).

\bibitem{Iskhakov2016b} T.Sh. Iskhakov, V.C. Usenko, R. Filip, M.V. Chekhova, and G. Leuchs, Low-noise macroscopic twin beams, {\it Phys. Rev.} A {\bf 93}, 043849 (2016).

\bibitem{Parigi2007} V. Parigi, A. Zavatta, M. Kim, and M. Bellini, Probing quantum commutation rules by addition and subtraction of single photons to/from a light field, {\it Science} {\bf 317}, 1890 (2007).

\bibitem{Zavatta2008} A. Zavatta, V. Parigi, M. S. Kim, and M. Bellini, Subtracting photons from arbitrary light fields: experimental test of coherent state invariance by single-photon annihilation, {\it New J. Phys.} {\bf 10}, 123006 (2008).

\bibitem{Fedorov2015} I. A. Fedorov, A. E. Ulanov, Y. V. Kurochkin, and A. I. Lvovsky, Quantum vampire: collapse-free action at a distance by the photon annihilation operator,  {\it Optica} {\bf 2}, 112 (2015).

\bibitem{Barnett2018}
S. M. Barnett, G. Ferenczi, C. R. Gilson, and F. C. Speirits, Statistics of photon-subtracted and photon-added states, {\it Phys. Rev.} A {\bf 98}, 013809 (2018).

\bibitem{Bogdanov2017} Y. I. Bogdanov, K. G. Katamadze, G. V. Avosopiants, L. V. Belinsky, N. A. Bogdanova, A. A. Kalinkin, and S. P. Kulik, Multiphoton subtracted thermal states: Description, preparation, and reconstruction, {\it Phys. Rev.} A {\bf 96}, 063803 (2017).

\bibitem{Katamadze2018} K. G. Katamadze, G. V. Avosopiants, Y. I. Bogdanov, and S. P. Kulik, How quantum is the quantum vampire effect?: testing with thermal light, {\it Optica} {\bf 5}, 723 (2018). 

\bibitem{Katamadze2019} K. G. Katamadze, E. V. Kovlakov, G. V. Avosopiants, and S. P. Kulik, Direct test of the quantum vampire's shadow absence with use of thermal light, {\it Opt. Lett.} {\bf 44}, 3286 (2019).

\bibitem{Katamadze2020} K. G. Katamadze, G. V. Avosopiants, N. A. Bogdanova, Yu. I. Bogdanov, and S. P. Kulik, Multimode thermal states with multiphoton subtraction: Study of the photon-number distribution in the selected subsystem, {\it Phys. Rev.} A {\bf 101},  (2020).

\bibitem{Enzian2021} G. Enzian, J. J. Price, L. Freisem, J. Nunn, J. Janousek,  B. C. Buchler, P. K. Lam, and M. R. Vanner, Single-phonon addition and subtraction to a mechanical thermal state, {\it Phys. Rev. Lett.} {\bf 126}, 033601 (2021). 

\bibitem{Vidrighin2016} M. D. Vidrighin, O. Dahlsten, M. Barbieri, M. S. Kim, V. Vedral, and I. A. Walmsley, Photonic Maxwell's demon, {\it Phys. Rev. Lett.} {\bf 116}, 050401 (2016).

\bibitem{Shu2017} A. Shu, J. Dai, and V. Scarani, Power of an optical Maxwell's demon in the presence of photon-number correlations, {\it Phys. Rev.} A {\bf 95}, 022123 (2017).

\bibitem{Hlousek2017} J. Hlou{\v s}ek, M. Je\v zek, and R. Filip, Work and information from thermal states after subtraction of energy quanta, {\it Sci. Rep.} {\bf 7}, 13046 (2017).

\bibitem{Hlousek2022} J. Hlou\v sek, T. Denzler, V. \v Svarc, M. Je\v zek, E. Lutz, and R. Filip, Experimental realization of a quantum photonic Maxwell demon, will be on Arxive.

\bibitem{Zanin2022}
G. L. Zanin, M. Antesberger, M. J. Jacquet, P. H. S. Ribeiro, L. A. Rozema, and P. Walther, Enhanced photonic Maxwell's demon with correlated baths, {\it Quantum} {\bf 6}, 810 (2022).

\bibitem{Spagnolo2022}
M. Spagnolo, J. Morris, S. Piacentini, M. Antesberger, F. Massa, A. Crespi, F. Ceccarelli, R. Osellame, and P. Walther, Experimental photonic quantum memristor, {\it Nat. Photon.} {\bf 16}, 318 (2022).

\bibitem{Leibfried2003} D. Leibfried, R. Blatt, C. Monroe, and D. Wineland, Quantum dynamics of single trapped ions, {\it Rev. Mod. Phys.} {\bf 75}, 281 (2003).

\bibitem{Chu2018} Y. Chu, P. Kharel, T. Yoon, L. Frunzio, P.T. Rakich and R.J. Schoelkopf, Creation and control of multi-phonon Fock states in a bulk acoustic-wave resonator, {\it Nature} {\bf 563}, 666 (2018).

\bibitem{Shore1993} B.W. Shore and P.L. Knight, The Jaynes-Cummings Model, {\it J. Mod. Opt.} {\bf 40}, 1195 (1993).

\bibitem{Fink2008} J. M. Fink, M. G{\"o}ppl, M. Baur, R. Bianchetti, P. J. Leek, A. Blais and A. Wallraff, Climbing the Jaynes-Cummings ladder and observing its nonlinearity in a cavity QED system, {\it Nature} {\bf 454}, 315-318 (2008).

\bibitem{Um2016} Mark Um, Junhua Zhang, Dingshun Lv, Yao Lu, Shuoming An, Jing-Ning Zhang, Hyunchul Nha, M. S. Kim and Kihwan Kim,  Phonon arithmetic in a trapped ion system, {\it Nature Communications} {\bf 7}, 11410 (2016).

\bibitem{Sukumar1981} C. V. Sukumar and B. Buck, Multi-phonon generalisation of the Jaynes-Cummings model, {\it Phys. Lett.} {\bf 83A}, 211 (1981).

\bibitem{Singh1982} S. Singh, Field statistics in some generalized JaynesCummings models, {\it Phys. Rev.} A {\bf 25}, 3206 (1982).

\bibitem{Villas-Boas2019} Celso J. Villas-Boas and Daniel Z. Rossatto, Multiphoton Jaynes-Cummings Model: Arbitrary Rotations in Fock Space and Quantum Filters, {\it Phys. Rev. Lett.} {\bf 122}, 123604 (2019).

\bibitem{McCormick2019} K.C. McCormick, J. Keller, S.C. Burd, D.J. Wineland, A.C. Wilson and D. Leibfried, Quantum-enhanced sensing of a single-ion mechanical oscillator, {\it Nature}  {\bf 572}, 86 (2019).

\bibitem{Scully1997} M.O. Scully and M.S. Zubairy, {\it Quantum Optics}, Cambridge University Press; 1st edition (1997).

\bibitem{Brune1987} M. Brune, J. M. Raimond, P. Goy, L. Davidovich and S. Haroche, Realization of a two-photon maser oscillator, {\it Phys. Rev. Lett.} {\bf 59}, 1899.

\bibitem{Garziano2015} L. Garziano, R. Stassi, V. Macr{\`i}, A. F. Kockum, S. Savasta and F. Nori, Multiphoton quantum Rabi oscillations in ultrastrong cavity QED, {\it Phys. Rev.} A {\bf 92}, 063830 (2015).

\bibitem{Ding2017a}
S. Ding, G. Maslennikov, R. Habl{\" u}tzel, H. Loh, and D. Matsukevich, Quantum parametric oscillator with trapped ions, {\it Phys. Rev. Lett.} {\bf 119}, 150404 (2017).

\bibitem{Ding2017b}
S. Ding, G. Maslennikov, R. Habl{\" u}tzel, and D. Matsukevich, Cross-Kerr nonlinearity for phonon counting, {\it Phys. Rev. Lett.} {\bf 119}, 193602 (2017).

\bibitem{Maslennikov2019}
G. Maslennikov, S. Ding, R. Habl{\" u}tzel, J. Gan, A. Roulet, S. Nimmrichter, J. Dai, V. Scarani, and D. Matsukevich, Quantum absorption refrigerator with trapped ions, {\it Nat. Commun.} {\bf 10}, 202 (2019).

\bibitem{Ding2018}
S. Ding, G. Maslennikov, R. Habl{\" u}tzel, and D. Matsukevich, Quantum simulation with a trilinear Hamiltonian, {\it Phys. Rev. Lett.} {\bf 121}, 130502 (2018).

\bibitem{Marek2018} P. Marek, J. Provazn{\'i}k and R. Filip, Loop-based subtraction of a single photon from a traveling beam of light, {\it Opt. Express} {\bf 26}, 29837 (2018).


\bibitem{Knight1982}
P. L. Knight and P. M. Radmore, Quantum revivals of a two-level system driven by chaotic radiation, {\it Phys. Lett} {\bf 90A} 342 (1982).

\end{thebibliography}

\appendix
\section{Optimal interaction times}
In semi-classical treatment $\bar{n}\gg 1$, the probability of a qubit in its ground state being excited by interacting with a thermal oscillator having a large mean phonon number $\bar{n}\gg 1 $ as can be approximated as \cite{Knight1982}
\begin{align}\label{eq:A1}
P_{\rm e}(t)\approx\lambda t\sqrt{\bar{n}}D(\lambda t \sqrt{\bar{n}})
\end{align}
where $\lambda$ is the coupling strength between the qubit and phonons, and $D(x)$ is Dawson's integral, defined by
\begin{align}\label{eq:A2}
D(x)=e^{-x^2}\int^{x}_0e^{x'^2}{\rm d}x'.
\end{align}
Before reaching the quantum-revival region, this semi-classical approximation agrees well with the calculation obtained from  rigorous quantum treatment. The probability $P_{\rm e}$ is, therefore, maximized when $\lambda t \sqrt{\bar{n}}\approx 1.502$. 

On the other hand, a thermal oscillator with small mean phonon number, $\bar{n}\ll 1$ dominantly occupies in its ground and the first excited states, $|0\rangle$ and $|1\rangle$. The population of all other motional states  becomes negligible compared to that of the two states. It is the phonon in the Fock state $\ket{1}$ that mainly excites a qubit. The probability of successful excitation of a qubit in this case then becomes
\begin{align}\label{eq:A3}
P_{\rm e}(t)=\sum^{\infty}_{m=0} p_m\sin^2(\lambda t \sqrt{m})
\approx \bar{n} \sin^2(\lambda t \sqrt{1})+\mathcal{O}(\bar{n}^2).
\end{align}
This imply that the maximum  $P_{\rm e}$ occurs when $\lambda t \approx \pi/2$. To compromise between these two extreme limits, $\bar{n}\gg 1$ and $\bar{n}\ll 1$, we then approximate the optimal interaction time as $\lambda t^{\rm op}_{0}\approx \pi/(2\sqrt{ \bar{n}+1})$, which approaches $\pi/2$ for small $\bar{n}$ and still well agrees with the semi-classical treatment for large $\bar{n}$. We note here that this approximation still hold true even in the intermediate limit of the average phonon number $\bar{n}$. For example, when $\bar{n}=2$, the approximation of $\lambda t^{\rm op}_{0}\approx \pi/(2\sqrt{\bar{n}+1}) $ is  differed from its actual value only by $\sim 1\%$.

For the semi-classical case, $\bar{n}\gg 1$, after several linear subtractions the population distribution of phonons is modified into a bell-shaped with its peak centered around its average phonon number $\langle \hat{n}\rangle$ and can, therefore, be qualitatively approximated as a Gaussian distribution as
\begin{align}\label{eq:A4}
p_m\sim \frac{1}{\sqrt{2\pi}\sigma}\exp\left(-\frac{(m-\langle\hat{n}\rangle)^2}{2\sigma^2}\right),
\end{align}
where $\sigma^2$ represents the variance of the distribution. The optimal interaction times for linear subtractions for the approximated Gaussian distribution is then of the form $\lambda t^{\rm op}_N\approx \pi/(2\sqrt{\langle \hat{n}\rangle +1})$, which is similar to the previous result. On the other hand, for this bell-shaped distribution, the probability associated with a successful nonlinear subtraction becomes
\begin{align}\label{eq:A5}
P^{\prime}_{\rm e}\approx \frac{1-e^{-2\lambda'^2 t^2\sigma^2}\cos(2\lambda^\prime t (\langle \hat{n}\rangle))}{2}.
\end{align}
With a sufficiently small variance $\sigma^2$, we therefore can qualitatively approximate the optimal interaction time for a nonlinear subtraction as $\lambda^\prime \tau^{\rm op}\approx \pi/2\langle \hat{n}\rangle$, as shown in Eq. \ref{eq:4.5}.
\section{Truncation}
The approximated optimal interaction times, presented previously, allow us to analyze how linear and nonlinear phonon subtractions modify the population distribution. Let us  first begin with linear subtractions. We particularly consider the first terms of Eqs. \ref{eq:2.6} and \ref{eq:2.8}
\begin{align}\label{eq:A6}
  p^{(N)}_m = p^{(N-1)}_{m+1}\sin^2(\lambda t^{\rm op}_{N-1}\sqrt{m+1})+{\rm second\ term},
\end{align}
which are the dominant terms of the equations. As $\lambda t^{\rm op}_{N-1}\approx\pi/(2\sqrt{\langle\hat{n}\rangle+1})$, it is straightforward to realize that after a linear subtraction the populations $p_m$ of the Fock states $\ket{m}$ with $m\sim4 k^2\langle \hat{n}\rangle$, where $k$ is an arbitrary integer, are  suppressed the most due to the small values of the sine function,  as
\begin{align}\label{eq:A7}
\sin^2(\lambda t^{\rm op}_{N-1}\sqrt{4k^2\langle \hat{n}\rangle+1})&\approx\sin^2\left(\frac{\pi}{2}\sqrt{\frac{4k^2\langle \hat{n}\rangle+1}{\langle \hat{n}\rangle+1}}\right)\\
&\sim\sin^2\left(k\pi\right)= 0.
\end{align}
We then can consider the first two most-suppressed points in the population distribution, $m=0$ and $m\sim 4\langle \hat{n}\rangle$, as the locations at which linear subtractions trim the phonon distribution. However, the population of a thermal phonon being in an energy level  $m> 4\langle\hat{n}\rangle$ is already relatively very small at the beginning. Especially in the semi-classical case $\bar{n}\gg 1$, the estimated probability of a thermal phonon populating in such high energy levels is already much less than unity as
\begin{align}\label{eq:A8}
P(m>4\bar{n})=\sum^{\infty}_{m>4\bar{n}}\frac{\bar{n}^m}{(\bar{n}+1)^{m+1}}\sim \exp(-4)= 0.018.
\end{align}
A ripple in the distribution originated from the sine function in Eq. \ref{eq:A6}, which in principle can occur in the region of $m>4\bar{n}$, therefore, becomes relatively insignificant and can be omitted as the population in such region is already very small. This means linear subtractions can truncate the tails of population distribution desirably as they do not create a noticeable ripple in the distribution.

\begin{figure}[h!]
\centering
\includegraphics[width=\linewidth]{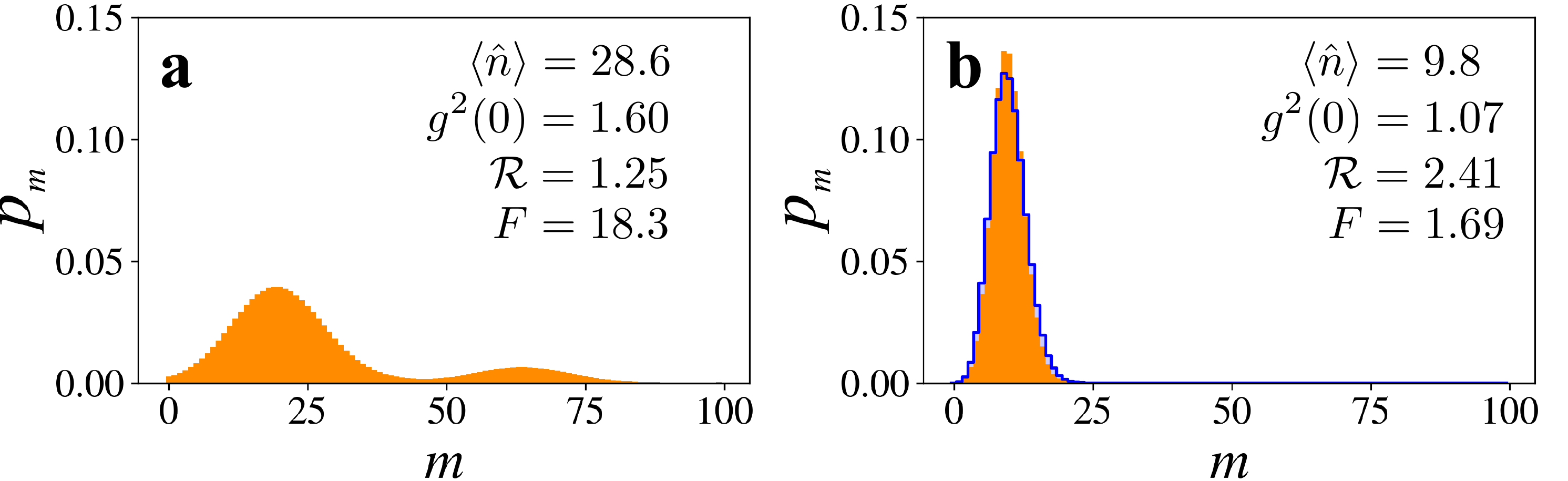}
\caption{({\bf a})The phonon population distribution obtained from three phonon subtractions, a linear subtraction followed by two consecutive nonlinear subtractions, for the initial thermal state with $\bar{n}=30$. We can clearly see the ripple of the distribution as the populations of the energy levels around $m\sim 45$ and $m\sim 80$ are greatly suppressed due to the nonlinear interaction. ({\bf b}) The distribution after eliminating the ripple in Figure \ref{fig:7}a using 8 linear subtractions and squeezing the main part of the distribution further by 4 nonlinear subtractions. Its statistics approaches Poissonian. A sub-Poissonian is prevented by the small, but finite population of very high energy Fock states.}\label{fig:7}
\end{figure}

For nonlinear subtractions, we can analyze its effect on the phonon distribution in a similar manner. The modification of the distribution due to a nonlinear subtraction is described by Eq. \ref{eq:4.6}. To focus on the first term, the dominant term of the equation, we then can rewrite this equation as
\begin{align}
p^{\rm f}_m=p^{\rm i}_{m+2}\sin^2\left(\lambda^{\prime} \tau^{\rm op}\sqrt{(m+2)(m+1)}\right)+{\rm second\ term}
\end{align}
For a bell-shaped distribution with $\langle \hat{n}\rangle\gg 1$ and the approximated optimal time$\lambda^\prime \tau^{\rm op}\approx \pi/2\langle \hat{n}\rangle$, the population in the energy levels around $n\sim 2 k\langle \hat{n}\rangle$, where again $k$ is an arbitrary integer, are suppressed the most. We then again select the first two most suppressed points, $m=0$ and $m\sim 2\langle\hat{n}\rangle$, to be the truncated points of the population distribution for this case. The oscillation in phonon population due to nonlinear subtractions can still apparently appear in the range $m> 2\langle \hat{n}\rangle$, if the population of phonons being in a Fock state higher than $2\langle \hat{n}\rangle$ is not yet sufficiently small beforehand. Nonlinear subtractions then can undesirably generate a ripple in the phonon population. To explicitly visualize it, let us consider the case when a nonlinear subtraction is performed after just a single linear subtraction. The population is not yet shaped properly by the linear subtraction. We, however, can somehow qualitatively approximated the optimal interaction time of the nonlinear subtraction as $\lambda^\prime \tau^{\rm op}\sim \pi/2
\bar{n}$ for $\bar{n}\gg 1$.  The nonlinear subtraction thus will trim the population around $m=0$ and $m\sim 2\bar{n}$. As the population of phonons in energy levels $m>2\bar{n}$ is not yet sufficiently small,  the oscillation in the phonon population then becomes apparent as shown in Fig \ref{fig:7}a. On the other hand, as the first highest peak noticeably becomes a bell-shaped, we may also exploit several linear subtractions to suppress the ripple and then further squeezes it with several nonlinear subtractions. The {\it main} peak can be even more squeezed than a Possionian as shown in figure \ref{fig:7}b. We note here that the shape of the distribution in the figure may resembles a sub-Possonian, but its second-order coherence function $g^{(2)}(0)$ implies otherwise as $g^{(2)}(0)>1$. This is because there is still a very small, but finite probability that very high energy Fock states are populated.

\end{document}